\documentclass[1p]{elsarticle}

\usepackage{lineno,hyperref}

\journal{Journal of Computational Physics}

\usepackage[utf8]{inputenc}
\usepackage[T1]{fontenc}
\usepackage[english]{babel}
\usepackage{algpseudocode}
\usepackage{amsmath}
\usepackage{amsfonts}
\usepackage{latexsym}
\usepackage{amssymb}
\usepackage{graphicx}
\usepackage{color}
\usepackage{dsfont}
\usepackage{psfrag}
\usepackage{bbm}
\usepackage{setspace}
\usepackage{float}
\usepackage{siunitx}
\usepackage{mathtools}

\usepackage{subcaption}
\usepackage{cleveref}
\usepackage{bm}
\usepackage[OMLmathsfit]{isomath}

\usepackage{pgfplots}
\usepackage{lineno}
\pgfplotsset{compat=newest} 
\pgfplotsset{plot coordinates/math parser=false} 
\newlength\figureheight 
\newlength\figurewidth

\newcommand{\vv}[1]{\bm{#1}}
\newcommand{\mc}[1]{\mathcal{#1}}
\newcommand{\p}[1]{\partial{#1}}
\newcommand{\mb}[1]{\mathbb{#1}}

\newcommand{\pn}{\partial_{\hat{\bm{n}}}} 
\newcommand{\pnp}{\partial_{\hat{\bm{n}}^\prime}}

\newcommand{\mat}[1]{\mathboldsans{#1}}
\newcommand{\diff}{\,\mathrm{d}}

\newcommand\restr[2]{{
		\left.\kern-\nulldelimiterspace 
		#1 
		\vphantom{|} 
		\right|_{#2} 
}}

\newcommand\rst[3]{{
		\left.\kern-\nulldelimiterspace 
		#1 
		\vphantom{|} 
		\right|_{#2}^{#3} 
}}

\newcommand{\T}{\mathrm{T}}

\biboptions{sort & compress}

\begin{document}
	\begin{frontmatter}

		
		
		\title{On the modeling of brain fibers in the EEG forward problem via a new family of wire integral equations}
		

\author[polito]{ Lyes Rahmouni } 
\author[polito]{Adrien Merlini }
\author{Axelle Pillain}
\author[polito]{Francesco P. Andriulli\corref{mycorrespondingauthor}}
\address[polito]{Department~of~Electronics~and~Telecommunications, Politecnico~di~Torino, Turin, Italy}


\cortext[mycorrespondingauthor]{Corresponding author}
\ead{francesco.andriulli@polito.it}

\begin{abstract}
Source localization based on electroencephalography (EEG) has become a widely used neuroimagining technique. However its precision has been shown to be very dependent on how accurately the brain, head and scalp can be electrically modeled within the so-called  forward problem. The construction of this  model is traditionally performed by leveraging Finite Element or Boundary Element Methods (FEM or BEM). Even though the latter is more computationally efficient thanks to the smaller interaction matrices it yields and near-linear solvers, it has traditionally been used on simpler models than the former. Indeed, while FEM models taking into account the different media anisotropies are widely available, BEM models have been limited to isotropic, piecewise homogeneous models. In this work we introduce a new BEM scheme taking into account the anisotropies  of the white matter. The boundary nature of the formulation allows for an efficient discretization and modelling of the fibrous nature of the white matter as one-dimensional basis functions, limiting the computational impact of their modelling. We compare our scheme against widely used formulations and establish its correctness in both canonical and realistic cases.
\end{abstract}

\begin{keyword}
			
Electroencephalography, EEG forward problem \sep Anisotropy \sep Integral equations \sep Boundary Element Method
			
			
\end{keyword}
\end{frontmatter}
	
\section{Introduction}

Electroencephalography (EEG) based source localization has gained an increasing popularity as a reliable neuroimaging modality in research and medical practice \cite{thut2009new, van1998technical, nemtsas2017source}. Using scalp measured potentials, various algorithms have been proposed for the retrieval of the location of the neuro-generators \cite{pascual1999review}. Many of these algorithms rely on an accurate solution of the associated forward problem which maps a given setting of sources and head model to the corresponding scalp potential. The complexity of the head geometry and its underlying conductivity, however, precludes the use of analytical methods and one has to adopt numerical approximations. With their renowned high accuracy and robustness, integral equations-based methods remain the preferred choice for researchers \cite{hedrich2017comparison, khosropanah2018fused}. In particular, the boundary element method (BEM) only requires the discretization of the boundaries, thus reducing the overall dimensionality. Moreover, given the smoothness of its underlying kernel, it is possible to augment BEM with fast algorithms such as the adaptive cross approximation (ACA) or the fast multipole method (FMM) \cite{greengard1998fast, ostrowski2006fast}, which further reduce its computational complexity. 
The three most widely employed BEM formulations for the EEG forward problem are the adjoint double layer (ADL), the double layer (DL) and the symmetric (SY) approaches \cite{stenroos2012bioelectromagnetic, adde2003symmetric, rahmouni2018conforming}. By leveraging on methods of layer potentials, these methods solve Poisson equation under the assumption of isotropic media \cite{kybic2005common}. The DL formulation is a direct approach in which the potential is obtained directly while the ADL formulation solves first for an auxiliary unknown before integrating it to obtain the electric potential. Differently from the two previous approaches, the SY formulation simultaneously involves two surface unknowns. Despite its larger system of equations, it has a block diagonal structure \cite{kybic2005common}. 
For more details of these methods, their relative merits and disadvantages, the reader is referred to \cite{pillain2016handling, adde2003symmetric, rahmouni2018conforming}.

Despite their advantages, BEM-based formulations are restricted to isotropic and piece-wise homogeneous problems. This is a significant limitation since white matter anisotropy has a considerable impact \cite{wolters2006influence, haueisen2002influence} on the accuracy of source localization procedures.
These early results have been obtained with differential based methods and entire volume discretization \cite{wolters2006influence,haueisen2002influence}, which is computationally expensive. More recently, integral techniques accounting for the white matter anisotropy have been introduced \cite{rahmouni2017two,pillain2016handling}; they do however also require discretization of the entire head volume.

The anisotropy of the white matter tissue arises from its underlying assembly of bundles of parallelly-oriented axon \cite{friman2006bayesian,lazar2003white}.  This suggests that the apparent inhomogeneous anisotropy is actually structured and may be expressed in terms of these axons’ fibers. 
This observation has been leveraged on in \cite{olivi2011handling} by replacing a single fiber by dipolar sources of constant magnitude. The forward problem was subsequently solved iteratively with the symmetric formulation. However, this work does not account for the coupling and interactions between different fibers which is essential for precise forward solution.

The work presented in this paper aims at extending the three main BEM (EEG) formulations to take into account the anisotropic and inhomogeneous conductivity of the white matter. This is achieved by a modelization of the white matter connectivity. Indeed, using diffusion weighted MRI (DW-MRI) it is possible to track axon fibers and reveal the underlying network of the white matter \cite{mars2011diffusion}.
One-dimensional basis functions are used for the modelization of the fibers which results in efficient and accurate forward solutions. 
As a byproduct, the new technique we present could further improve the recently introduced approaches exploiting the brain connectivity patterns in source estimation \cite{hammond2013cortical,hammond2012incorporating}. Some preliminary results have been presented in \cite{rahmouni2017integral}. Several numerical experiments validate the new schemes in canonical and realistic settings.

{\color{black} The reader should note that 1D formulations have been extensively studied in the context of high frequency electromagnetic modeling of wire-like structures \cite{papakanellosExtendedThinWireKernel2016,mohanAccurateModelingCylindrical2006,wilton2006evaluation}, although those schemes, for perfect electrically conducting wires, are only mildly related to the ones presented here.}
	
The paper is organized as follow: the notations is set and some background is recalled in \Cref{sec:background}; the new equations and their discretizations are then derived in \Cref{sec:integral_equation} and \Cref{sec:discretization}, respectively. The new schemes are validated with various simulations and tests in \Cref{sec:numerical_results} before closing with conclusions in \Cref{sec:conclusion}.

	
	\section{Background and notations} \label{sec:background}

\begin{figure}
		\centering
		\includegraphics[width=0.35\linewidth]{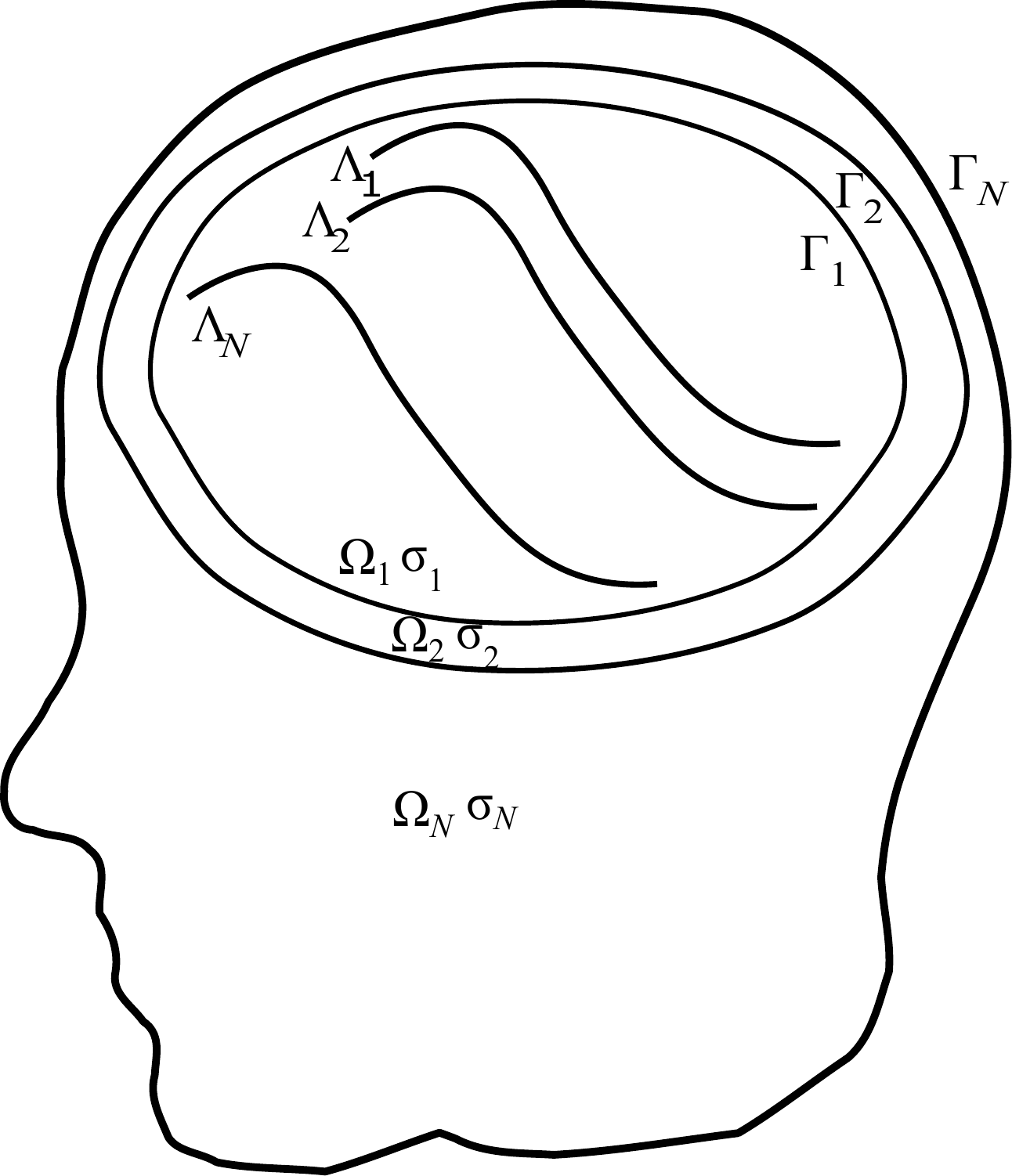}
		\caption{Volume conductor with nested geometry. }
		\label{fig:NestedGeo}
	\end{figure}

	Consider an electric volume  current density $\vv J$ residing in a conducting medium $\Omega \subset \mathbb R^3$ composed of $N$ nested, piecewise homogeneous sub-regions $\Omega_\text{i}$ such that $\Omega  = \bigcup\nolimits_{i = 1}^N {{\Omega _\text{i}}}$ with $\Omega_1$ being the innermost layer. Each sub-region is associated with  an isotropic conductivity $\sigma_i$ and delimited by the Lipschitz surface $\Gamma_i$ with $\Gamma  = \bigcup\nolimits_{i = 1}^N {{\Gamma _\text{i}}}$ and $\Gamma_i \cup \Gamma_{i-1}= \partial \Omega_i$ (\Cref{fig:NestedGeo}).
	%
	
	%
	Let $\Lambda_\text{j}$ be a curve modeling a bundle of parallel white matter fibers and $\Lambda=\bigcup_{j = 1}^{N_\text{f}} {{\Lambda _\text{j}} }$  their union. These fibers assume a thin cylindrical shape of circular section $a$ \cite{werring1999diffusion, chen2013detecting} and potentially contain junctions; they give rise to a tensorial conductivity profile in which the conductivity along the fiber $\sigma _f$ is different from the  conductivity in the transversal direction $\sigma_1$. In the quasi-static regime, the electric potential $\phi$ is related to the current density $\vv J$  via Poisson's equation
	\begin{equation}
	    \label{eq:Poisson}	
	    \nabla \cdot \left( \underline{\underline \sigma } (\vv r)  \nabla \phi(\vv r) \right) = \nabla \cdot \vv{J}(\vv{r})\,, \quad \vv r \in \Omega\,,
	\end{equation}
	where the local electric conductivity $\underline{\underline \sigma }$ is  described by a $3 \times 3$ symmetric tensor \cite{wolters2003influence}.  
	For the sake of simplicity in explaining our new method, we neglect the anisotropy of the skull and thus, given that the neuron's fibers are only present in the innermost region $\Omega_1$, \cref{eq:Poisson} can be rewritten as
	\begin{align}
	    \label{eq:Poisson1}
	    \nabla \cdot \left( \underline{\underline \sigma } (\vv r)\nabla \phi (\vv r)\right) &= \nabla \cdot\vv J (\vv r) \,,& \vv r \in {\Omega_1},\\
	    \label{eq:Poisson2}
	    {\sigma_i} \upDelta \phi (\vv r) &= 0 \,,&\vv r \in {\Omega _i},\,i = 2,\dots,N .
	\end{align}
where it was assumed that the current sources are present only in the innermost layer that corresponds to the brain. Two transmission conditions are associated with each of these equations: (i) a Dirichlet condition that enforces the continuity of the electric potential across  interfaces and (ii) a Neumann condition that enforces the continuity of the electric current flux, i.e.
	\begin{align}
	    \left[ \phi  \right]_j &= 0 \quad\quad \text{on~} \Gamma_j, \text{~for ~} j=1 \dots N \label{eq:BC1},\\
	    \left[ \sigma \pn \phi  \right]_j &= 0 \quad\quad \text{on~} \Gamma_j, \text{~for ~} j=1 \dots N \label{eq:BC2},
	\end{align}
	where the bracket notation $[g]_j$ denotes the jump of a function $g$ across  $\Gamma_j$ and $\pn g = \hat{\bm{n}}  \cdot \nabla g$ with $\hat{\bm{n}}  = \hat{\bm{n}} (\vv r)$ is the unit vector normal to $\Gamma_j$  pointing outward of $\Omega_j$. Note that the fibers do not come into contact with the inner surface.
	In the context of the EEG forward problem, brain sources are commonly modeled as current dipoles \cite{de1988mathematical, sarvas1987basic}, i.e. 
	\begin{equation}
    	\vv{J}(\vv r)=\vv P \delta(\vv r-\vv r_0),
	\end{equation}
	in which $\vv P$ and $\delta$ respectively denote the dipole moment and the Dirac delta function. The electric potential induced by this elementary source in an infinite homogeneous domain of conductivity $\sigma_{1}$ reads
	\begin{equation}
	    {{{v}}_\text{dip}}(\vv{r}) = \frac{{\vv{P}\cdot(\vv{r} - \vv{r_0})}}{{4\pi \sigma_1 {{\left| {\vv{r} - \vv{r_0}} \right|}^3}}}\,.
	\end{equation}
	
	\section{Integral equation based formulations} \label{sec:integral_equation}
	
By transforming Poisson equation into an integral equation, conventional BEM formulations (SL, DL and SY), have been particularly attractive as they offer computational savings in comparison with other alternatives. In general, reformulating a partial differential equation as an integral expression requires knowledge of its fundamental solution. When considering the anisotropy of the white matter however, \cref{eq:Poisson1} involves position and orientation dependent tensors for which the corresponding fundamental solution does not exist in closed form, for general geometries. \Cref{eq:Poisson1} should then be recast into an equivalent one at a reduced dimensionality by extracting the Laplacian operators and using Green's identities. In particular, this choice not only allows for a unified treatment of \cref{eq:Poisson1,eq:Poisson2}, but also reduces the effect of the anisotropy to an one-dimensional apparent volume current density along the fibers. Consequently, the framework of standard BEM formulations can be extended to handle the anisotropy of the white matter. To that end, \cref{eq:Poisson1} can be re-expressed as 
	\begin{equation}
	    \label{eq:Poisson3}
	    \upDelta \phi (\vv r) = \nabla \cdot \left( \frac{\vv J (\vv r)}{\sigma_1}  +  \underline{\underline \kappa } (\vv r)\vv{\mathbb{J}} _f (\vv r)\right)\,,  \quad \quad \vv r \in {\Omega _1},
	\end{equation}
	where $\vv{\mathbb{J}} _f = \underline{\underline \sigma} \nabla \phi$ is the apparent volume current density along the fibers and $ \underline{\underline \kappa}$ is the conductivity contrast defined as
		\begin{equation}
        \underline{\underline \kappa} \left(\vv r\right) = \left(\underline{\underline \sigma}^{-1}(\vv r)  - {\sigma _1}^{-1} \underline{\underline I }  \right)\,,
    \end{equation}
	in which $\underline{\underline I }$ is the identity tensor. Note that $\underline{\underline \kappa}$ is zero everywhere except on the fibers where it has the form 
	\begin{equation}
	    \underline{\underline \kappa}(\vv r) = (\sigma _f^{-1} - \sigma _1^{-1}) \hat{\vv t}(\vv r) \, \hat{ \vv t}^\T(\vv r)
	\end{equation}
	 with $\hat {\vv t}(\vv r)$ being the  unit vector tangential to $\Lambda_j$. 
We remind the reader that it was assumed that the conductivity of the fibers is ${\sigma _1}$ transversally and $\sigma _f$ longitudinally. 
    In order to derive an integral representation for the potential and its flux using  Green's second identity, the following well known operators are introduced:
	\begin{itemize}
\item the single layer operator
	\begin{equation}
	    (\mc S \restr{u}{\Gamma_i})(\vv r) = \int_{\Gamma_i} G(\vv r-\vv r') u(\vv r') \diff s(\vv r')\,, \quad \vv r \in \Omega \,, \label{eq:OpS}
	\end{equation}
\item the double layer operator
	\begin{equation}
	    (\mc D\restr{u}{\Gamma_i})(\vv r) = \int_{\Gamma_i} \pnp G(\vv r-\vv r') u(\vv r') \diff s(\vv r')\,, \quad \vv r \in \Omega \,, \label{eq:DL}
	\end{equation}

\item the adjoint double layer operator
	\begin{equation}
	    (\mc K\restr{u}{\Gamma_i})(\vv r) = \int_{\Gamma_i} \pn G(\vv r-\vv r') u(\vv r') \diff s(\vv r')\,, \quad \vv r \in \Gamma_j \,,\label{eq:AdjDL}
	\end{equation}
\item the hypersingular operator
	\begin{equation}
	    (\mc N\restr{u}{\Gamma_i})(\vv r) = \int_{\Gamma_i} \pn \pnp G(\vv r-\vv r') u(\vv r) \diff s(\vv r')\,, \quad \vv r \in \Gamma_j  \,, \label{eq:OpN}
	\end{equation}
\end{itemize}
where
	\begin{equation}
	    G(\vv r-\vv r') = \frac{1}{{4\pi \left| {\vv r - \vv r'} \right|}}
	\end{equation}
	is the fundamental solution associated with the Laplacian.
In addition, a new operator is introduced to handle the fiber contributions 
	\begin{equation}
	    (\mc V \vv u)(\vv r) = \int_\Lambda g(\vv r-\vv r') \nabla \cdot (\underline{\underline{\kappa}} \vv u(\vv r')) \diff l( \vv r')\,, \quad \vv r \in \Omega\,,
	\end{equation}
where the associated wire kernel is defined as
\begin{equation}
	g(\vv r-\vv r') = \int_0^a {\int_0^{2\pi} G(\vv r-\vv r') \rho' \diff \rho' \diff \theta'  }\,,
\end{equation}
and $\rho$ and $\theta$ are the usual polar coordinates in the fiber's transverse plane.
	
The starting point of our development is Green's second identity, which states that:

	\begin{equation}
	    \label{eq:Green1_Vol1}
	    (\mc D \restr {\phi} {\p\Omega_i}) (\vv r)- (\mc S \restr {\xi} {\p\Omega_i}) (\vv r) = \int_{\Omega_i} \phi(\vv r)\upDelta G(\vv r-\vv r') - G(\vv r-\vv r')\upDelta \phi(\vv r) \quad \vv r \in \Omega_i\setminus \Gamma_i
	\end{equation}
	where  $\xi=\pn \phi$ is the derivative of the potential in the normal direction. Using \cref{eq:Poisson2,eq:Poisson3} and the property of the fundamental solution, \cref{eq:Green1_Vol1} reduces to the following
	\begin{multline}
	    \label{eq:Green1_Vol2}
	    (\mc D \restr {\phi} {\p\Omega_i}) (\vv r)- (\mc S \rst {\xi} {\p\Omega_i}{\pm}) (\vv r) =\\ \phi (\vv r)  + \left\{ \begin{array}{cl}
	    v_\text{dip} (\vv r)-
	    \sum\limits_{k=1}^{{N_\text{f}}} {\mc V \vv{\mathbb{J}}_{\Lambda_\text{k}} }  (\vv r) & \vv r \in { \Omega _i}\setminus \Gamma_i,\,i = 1\,, \\
	    0 & \vv r \in {\Omega _i}\setminus \Gamma_i,\,i = 2 \dots N\,,
	\end{array} \right.
	\end{multline}	
Taking the limit $\vv r \rightarrow \partial \Omega$, the following integral representation for the electric potential is derived	
	\begin{multline}
	    \label{eq:Green1}
	    (\mc D \restr {\phi} {\p\Omega_i}) (\vv r)- (\mc S \rst {\xi} {\p\Omega_i}{\pm}) (\vv r) =\\ - \frac{1}{2}\restr {\phi} {\p\Omega_i} (\vv r)  + \left\{ \begin{array}{cl}
	    v_\text{dip} (\vv r)-
	    \sum\limits_{k=1}^{{N_\text{f}}} {\mc V \vv{\mathbb{J}}_{\Lambda_\text{k}} }  (\vv r) & \vv r \in {\partial \Omega _i},\,i = 1\,,\\
	    0 & \vv r \in {\partial\Omega _i},\,i = 2 \dots N\,,
	\end{array} \right.
	\end{multline}
 By differentiating \cref{eq:Green1_Vol2} with respect to $\vv r$ in the direction normal to the boundary, an integral representation for the potential flux can be obtained
	\begin{multline}
	\label{eq:Green2}
	(\mc N \restr {\phi} {\p\Omega_i}) (\vv r)- (\mc K \rst {\xi} {\p\Omega_i}{\pm}) (\vv r) = \\ -\frac{1}{2} \rst {\xi} {\p\Omega_i}{\pm} (\vv r)  + \left\{ \begin{array}{cl}
	v_\text{s} (\vv r)-
	\sum\limits_{k=1}^{{N_\text{f}}} {\mc W \vv{\mathbb{J}}_{\Lambda_\text{k}} } (\vv r) & \vv r \in {\partial \Omega _i},\,i = 1\,,\\
	0 & \vv r \in {\partial \Omega _i},\,i = 2 \dots N\,,
	\end{array} \right.
	\end{multline}
	where
	\begin{equation}
	\label{eq:D_strF}
	(\mc W \vv u)(\vv r) = \int_\Lambda \pn g(\vv r-\vv r') \nabla \cdot \left( \underline{\underline{\kappa}} \vv u(\vv r')\right) \diff l(\vv r')\
	\end{equation}
	and
	\begin{equation}
	{ v_\text{s}}(\vv r) = \pn {v_\text{dip}}(\vv r).
	\end{equation}
	
	It is worth noting that \cref{eq:Green1,eq:Green2} are written for a normal vector pointing outward. A consistent change of signs should be made when the normals are pointing inward, which is the case for $\Gamma_{i-1}$.  In the inner most layer $\Omega_1$ the last term of the right-hand side in \cref{eq:Green1,eq:Green2} represents the effect of the fibers; it describes the local anisotropic conductivity. 

	\subsection{Double layer-Wire formulation}
	
	\Cref{eq:Green1,eq:Green2} have two surface unknowns, one of which could be discarded by using the boundary conditions \cref{eq:BC1,eq:BC2}; depending on the variable discarded two different formulations can be obtained. The double layer-wire formulation is obtained if the surface electric potential $\phi(\vv r)$ is the remaining unknown. This formulation can be derived after multiplying \cref{eq:Green1} with the local conductivity and summing the contribution of all the regions $\Omega_i$
	\begin{multline}
	\label{eq:DLF1}
	\sigma_1 v_\text{dip} (\vv r)- \sigma_1
	\sum\limits_{k=1}^{{N_\text{f}}} {\mathcal V \vv{\mathbb{J}}_{\Lambda_\text{k}} } (\vv r) = \frac{{{\sigma _j} + {\sigma _{j + 1}}}}{2}  \phi(\vv r) - \sum\limits_{i = 1}^N {\left( {{\sigma _{i + 1}} - {\sigma _i}} \right)(\mc D{ \restr{\phi} {\Gamma_{i}}}})(\vv r), \\ \vv r \in \Gamma_j, \, j=1 \dots N ,
	\end{multline}
	where the $\mc S$ operator term cancels out by enforcing the transmission condition \eqref{eq:BC2}.
	
	\Cref{eq:DLF1} simultaneously involves the surface potential $\phi$ and the current density $\vv{\mathbb{J}}$ as unknowns, and therefore needs to be complemented with a second equation. The second equation is obtained by applying the gradient operator to \cref{eq:DLF1}
	\begin{multline}
	    \label{eq:DLF2}
	    \sigma_1 {\nabla v_\text{dip}}(\vv r) - \sigma_1
	    \nabla \sum\limits_{k=1}^{{N_\text{f}}} {\mathcal V \vv{\mathbb{J}}_{\Lambda_\text{k}} }  (\vv r)= \sigma_{1} \underline{\underline \sigma }^{ - 1}_{\Lambda_n} \vv{\mathbb{J}}_{\Lambda_\text{n}}   (\vv r) - \nabla \sum\limits_{i = 1}^N {\left( {{\sigma _{i + 1}} - {\sigma _i}} \right)({\mc D} \restr{\phi} {\Gamma_i}})(\vv r), \\ \vv r \in \Lambda_n, \, n=1 \dots N_\text{f} .
	\end{multline}
	Combining \cref{eq:DLF2,eq:DLF1} constitutes the first new formulation and will be referred to as the double layer-wire (\textit{DLW}) formulation. 
	
	\subsection{Single layer-Wire formulation}
	
	Differently from the DLW that is formulated in terms of surface potentials, the single layer-Wire (SLW) formulation is derived from \cref{eq:Green2} and solves for the jump of the potential's normal derivative across an interface. Thus, applying \cref{eq:Green2} to each region $\Omega_j$ and summing up their contributions yields
	\begin{multline}
	\label{eq:SLF0}
	v_\text{s}(\vv r)-
	\sum\limits_{k=1}^{{N_\text{f}}} {\mc W {\mb{J}_{\Lambda_\text{k}}}} (\vv r)= \frac{1}{2}(\rst {\xi} {\Gamma_j}{-}+\rst {\xi} {\Gamma_j}{+})  (\vv r) - \sum\limits_{i = 1}^N {\mc K} {(\rst {\xi} {\Gamma_i}{-}-\rst {\xi} {\Gamma_i}{+})} (\vv r), \quad \vv r \in \Gamma_j,\, j=1 \dots N\,,
	\end{multline}
	where the $\mc N$ operator term cancels out by enforcing the transmission condition \eqref{eq:BC1}.
	After introducing
	\begin{equation}
	q_{\Gamma_j}=\rst {\xi} {\Gamma_j}{-}-\rst {\xi} {\Gamma_j}{+}=\left( {\frac{{{\sigma _{j + 1}} - {\sigma _j}}}{{{\sigma _{j + 1}}}}} \right)\rst {\xi} {\Gamma_j}{-}\,,
	\end{equation}
	the difference between normal derivatives can be expressed as
	\begin{equation}
	\label{eq:diff_norm}
	\rst {\xi} {\Gamma_j}{-} +\rst {\xi} {\Gamma_j}{+}=\left( {\frac{{\sigma _{j + 1}} + {\sigma _j}}{\sigma _{j + 1}-\sigma _{j }}} \right) \restr{q} {\Gamma_j}\,.
	\end{equation}
	Substituting back \cref{eq:diff_norm} in \cref{eq:SLF0} forms the single layer formulation 
	\begin{equation}
	\label{eq:SL1}
	v_\text{s} (\vv r)-
	\sum\limits_{k=1}^{{N_\text{f}}} {\mc W {\mb{J}_{\Lambda_\text{k}}}} (\vv r)= \frac{{{\sigma _j} + {\sigma _{j + 1}}}}{{2({\sigma _{j + 1}} - {\sigma _j})}} \restr{q} {\Gamma_j}(\vv r) - \sum\limits_{i = 1}^N (\mc K \restr{q} {\Gamma_i})(\vv r), \quad \vv r \in \Gamma_j, j=1 \dots N\,.
	\end{equation}
	Similarly to the DLW \cref{eq:SL1} exhibits two unknowns and needs to be complemented. The complementary equation will be derived from \cref{eq:Green1} by summing the contributions of all regions
	\begin{equation}
	\label{eq:SL2}
	\phi(\vv r) = {v_\text{dip}} (\vv r)- \sum\limits_{k=1}^{{N_\text{f}}} {\mc V \vv{\mathbb{J}}_{\Lambda_\text{k}} } (\vv r)+\sum\limits_{i = 1}^N (\mc S \restr{q}{\Gamma_i})(\vv r)\,,
	\end{equation}
	where $\mc D$ vanishes due to condition \cref{eq:BC1}.
	A current equation is obtained by applying the gradient operator to \cref{eq:SL2}
	\begin{equation}
	\label{eq:SL3}
	{\nabla v_\text{dip}} (\vv r)-
	\nabla \sum\limits_{k=1}^{{N_\text{f}}} {\mathcal V \vv{\mathbb{J}}_{\Lambda_\text{k}} } (\vv r)=\underline{\underline \sigma } ^{ - 1}_{\Lambda_n} \vv{\mathbb{J}}_{\Lambda_\text{n}} (\vv r) - \nabla \sum\limits_{i = 1}^N ({\mc S \restr{q} {\Gamma_i}})(\vv r), \quad \vv r \in \Lambda_\text{n}, n=1 \dots N_\text{f}\,.
	\end{equation}
	Subsequently to finding $q$, the electric potential can be computed via \cref{eq:SL2}.
	
	\subsection{Symmetric-Wire formulation}
	
	The symmetric formulation leverages on a combination of \cref{eq:Green1} and \cref{eq:Green2} applied, in contrast with the two previous formulations, to adjacent regions only. Summing these contributions yields
	\begin{multline}
	\label{eq:symPartial1}
	(\mc D \restr {\phi} {\p\Omega_{i-1} }) (\vv r)
	-(\mc D \restr {\phi} {\p\Omega_i }) (\vv r) 
	-(\mc S \rst {\xi} {\p\Omega_{i-1}}{\pm}) (\vv r) 
	+(\mc S \rst {\xi} {\p\Omega_{i}}{\pm}) (\vv r)
	\\=   \left\{ \begin{array}{cl}
	-v_\text{dip} (\vv r)
	+\sum\limits_{k=1}^{{N_\text{f}}} {\mc V \vv{\mathbb{J}}_{\Lambda_\text{k}} }  (\vv r) & \vv r \in {\partial \Omega _i},\,i = 1,\\
	0 & \vv r \in {\partial\Omega _i},\,i = 2 \dots N.
	\end{array} \right.
	\end{multline}
	The current flux $\restr{d}{\Gamma_i} = \sigma_i \pn \rst{\xi}{\Gamma_i}{-}= \sigma_{i+1} \pn \rst{\xi}{\Gamma_i}{+}$ (by virtue of condition \cref{eq:BC2}), can be substituted in \cref{eq:symPartial1}
	\begin{multline}
	\label{eq:Sym1}
	(\mc D \restr {\phi} {\p\Omega_{i-1} }) (\vv r)
	-(\mc D \restr {\phi} {\p\Omega_i }) (\vv r) 
	-\sigma_{\Omega_{i-1}}^{-1}(\mc S \restr {d} {\p\Omega_{i-1}}) (\vv r) 
	+\sigma_{\Omega_{i}}^{-1} (\mc S \restr {d} {\p\Omega_{i}}) (\vv r)
	\\=   \left\{ \begin{array}{cl}
	-v_\text{dip} (\vv r)
	+\sum\limits_{k=1}^{{N_\text{f}}} {\mc V \vv{\mathbb{J}}_{\Lambda_\text{k}} }  (\vv r) & \vv r \in {\partial \Omega _i},\,i = 1,\\
	0 & \vv r \in {\partial\Omega _i},\,i = 2 \dots N .
	\end{array} \right.
	\end{multline}
This expression constitutes the first equation of the symmetric formulation. It has three unknowns, the surface potential, the normal component of the surface current density and the fibers current density. Therefore, two other equations are needed. In order to derive a second equation, \cref{eq:Sym1} is multiplied by the local conductivity and applied to adjacent domains, yielding
	\begin{multline}
	\label{eq:Sym2}
	-\sigma_{\Omega_{i-1}}(\mc N \restr {\phi} {\p\Omega_{i-1}}) (\vv r)
	+\sigma_{\Omega_{i}}(\mc N \restr {\phi} {\p\Omega_{i}}) (\vv r)
	+ (\mc K \restr {d} {\p\Omega_{i-1} }) (\vv r) 
	- (\mc K \restr {d} {\p\Omega_{i} }) (\vv r)\\
	=  \left\{ \begin{array}{cl}
	-\sigma_{1} v_\text{s} (\vv r)
	+\sigma_{1}\sum\limits_{k=1}^{{N_\text{f}}} {\mc W \vv{\mathbb{J}}_{\Lambda_\text{k}} } (\vv r) & \vv r \in {\partial \Omega _i},\,i = 1,\\
	0 & \vv r \in {\partial \Omega _i},\,i = 2 \dots N.
	\end{array} \right.
	\end{multline}
	For the third equation, the gradient operator of \cref{eq:Green1} is applied to the innermost layer, which leads to the current equation
	\begin{multline}
	\label{eq:Sym3}
	{\nabla v_\text{dip}} (\vv r)-
	\nabla \sum\limits_{k=1}^{{N_\text{f}}} {\mathcal V \vv{\mathbb{J}}_{\Lambda_\text{k}} } (\vv r)=\underline{\underline \sigma } ^{ - 1}_{\Lambda_n} \vv{\mathbb{J}}_{\Lambda_\text{n}} (\vv r) - \nabla (\mc D \restr{\phi}{\Gamma_1})(\vv r)+ \sigma_1^{-1} \nabla (\mc S \restr{d} {\Gamma_1})(\vv r),\\ \vv r \in \Lambda,\, n=1 \dots N_\text{f}.
	\end{multline}
	Note that the quantities restricted to non-existing surfaces i.e. $\Gamma_0$ and $\Gamma_{N+1}$ are set to zero and that on the outermost layer  $\restr{d}{\Gamma_{N}}$ is identically Zero.
    This formulation requires the solution of two surface equations out of which the surface unknowns interact with only their immediate neighbors. This will give rise to a block diagonal matrix, thus reducing the apparently higher computational cost.
	
	\section{Discretization} \label{sec:discretization}
	
\begin{figure}
		\centering
		\begin{subfigure}{.5\textwidth}
			\centering
			\includegraphics[trim=3.5cm 0.5cm 2.5cm 6cm, clip=true, totalheight=0.12\textheight]{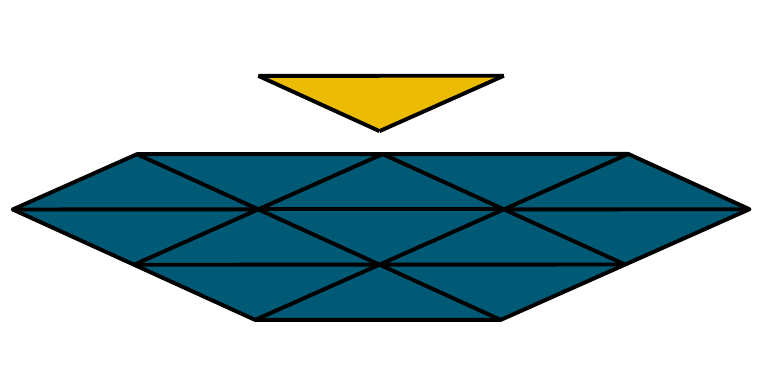}
			\subcaption{ \label{subfig:patch_bf}}
		\end{subfigure}%
		\begin{subfigure}{.5\textwidth}
			\centering
			\includegraphics[trim=3.1cm 2cm 3.0cm 2.8cm, clip=true, totalheight=0.14\textheight]{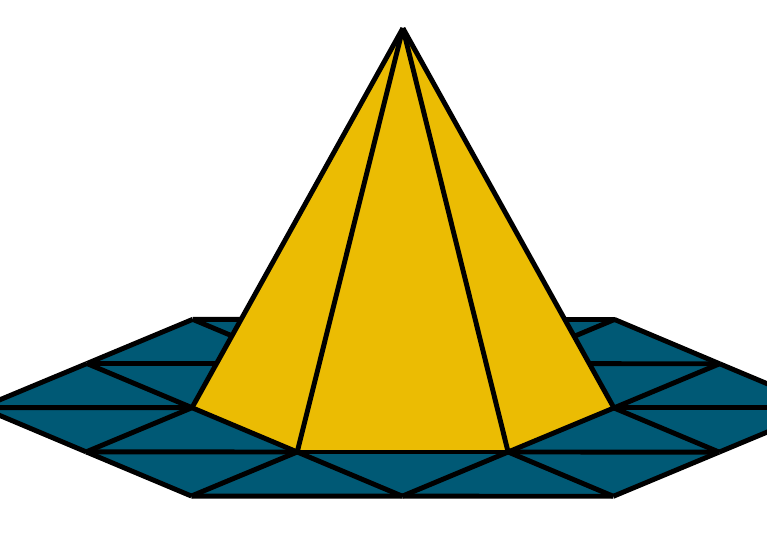}
			\subcaption{ \label{subfig:pyramid_bf}}
		\end{subfigure}
		\caption{Illustration of (\subref{subfig:patch_bf}) the patch (\subref{subfig:pyramid_bf}) the pyramid basis functions respectively defined in \cref{eq:Patch_Basis_Function} and \cref{eq:Pyramid_Basis_Function}.}
		\label{fig:Surf_BF}
	\end{figure}

	The numerical solution of the presented equations is achieved following a Galerkin approach. In this respect, the different head surfaces $\Gamma_{i}$ are tessellated into triangular meshes and the fibers $\Lambda_j$ into cylindrical segments. On these finite elements, each unknown $S(\vv r)$ is approximated by a linear combination of the $N_x$ basis functions $\{ x_i\}$
	\begin{equation}
	    S(\vv r) \approx \sum_{i = 1}^{N_x} {a_i x_i(\vv r)}\,,
	\end{equation}
where $a_i = \left< S(\vv r), x_i(\vv r)\right>$. In order to obtain a square linear system, the discretized equations are then tested with an appropriate set of functions of same cardinality as the set of basis functions. The choice of these finite elements is not arbitrary and must be in accordance with the operators' mapping properties i.e. the basis functions should span the domain of the operator and the testing functions should span the dual of its range \cite{rahmouni2018conforming, steinbach2007numerical}. The functions used to discretize the different unknowns must be capable of satisfying their different physical properties, for instance the discretization of the current density should not permit the existence of jumps.
In this paper we considered patch $\{\varphi_n(\vv r)\}$ and pyramid $\{\psi_n(\vv r)\}$ functions to expand the surface unknowns $\phi$, $q$ and $d$ depending of the formulation and hat basis functions $\{\vv \lambda_n(\vv r)\}$ to expand the current density $\mathbb{J}(\vv r)$.
The pyramid and patch basis functions $\psi_n(\vv r)$ are respectively expressed as
\begin{align}
    {\varphi _n}(\vv r) &=
\begin{dcases}
     1 & \text{if } \vv r \in T\text{r}_n\,,\\
     0 & \text{otherwise,}
\end{dcases} \label{eq:Patch_Basis_Function}\\
 \intertext{and,}
    {\psi_n}(\vv r) &=
\begin{dcases}
\frac{{\left| {\left( {\vv {{r_j}}  - \vv {{r_i}} } \right)\,\, \times \,\,\left( {\vv r  - \vv {{r_i}} } \right)} \right|}}{{\left| {\left( {\vv {{r_j}}  - \vv {{r_i}} } \right)\,\, \times \,\,\left( {\vv {{r_n}}  - \vv {{r_i}} } \right)} \right|}} & n \ne i \ne j \quad \text{if } \vv r \in T\text{r}_n\,,\\
	0
\end{dcases} \label{eq:Pyramid_Basis_Function}
\end{align}
where $\vv {r_{n}}, \vv {r_{i}}, \vv {r_{j}} $ are the position vectors of the vertices constituting the triangle $T\text{r}_n$. \Cref{subfig:pyramid_bf,subfig:patch_bf} presents the schematic definitions of these basis functions.

\begin{figure}
		\centering
		\includegraphics[width=1\textwidth]{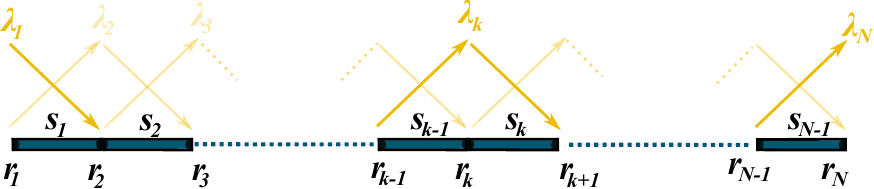}
		\caption{Illustration of the hat basis functions, as defined in \cref{eq:Hat_Basis_Function}.}
		\label{fig:hatbasisfunction}
	\end{figure}

	The current density is expanded with oriented hat functions whose support are the cylindrical segments
	$s_k = (\vv {r_k};\, \vv {r_{k+1}})$ and $s_{k+1} = (\vv {r_{k+1}};\, \vv {r_{k+2}})$ (\Cref{fig:hatbasisfunction}) and defined as
	\begin{equation}
    	{\vv \lambda _k}(\vv r)  =
\begin{dcases}
		\frac{  {\vv r - \vv {r_{k-1}}} }{{\left| \vv r_k - \vv {r_{k-1}} \right|}} & \text{if } \vv r \in {s_{k - 1}}\,,\\
		\frac{  \vv {r_{k+1}} - \vv r }{{\left| \vv r_k - \vv {r_{k+1}} \right|}} & \text{if } \vv r \in {s_{k}}\,,\\
		\vv 0 & \text{otherwise.}
\end{dcases} 
\end{equation}  \label{eq:Hat_Basis_Function}
	It should be noted that the hat basis functions are continuous and thus automatically enforce the jump condition of the current density.

	\subsection{Discretization of the double layer-wire formulation}
	
	In \cref{eq:DLF1,eq:DLF2}, the surface potential $\phi$ is discretized with pyramid basis functions and the current density $\vv{\mathbb{J}}$ is discretized with hat basis functions. \Cref{eq:DLF1,eq:DLF2} are then tested with pyramid and hat functions respectively. This gives rise to the following matrix system
	\begin{equation}
	\left[ {\begin{array}{*{20}{c}}
		{ \mat G_{\Lambda _m}^v + \mat V_{{\Lambda _m}{\Lambda _n}}^v}&\vline& {\mat D_{{\Gamma _n}{\Lambda _m}}^v}\\
		\\
		{\mat V_{{\Lambda _m}{\Gamma _n}}^s}&\vline& {\mat G_{\Gamma _n}^s + \mat D_{{\Gamma _n}{\Gamma _m}}^s}
		\end{array}} \right]\left[ {\begin{array}{*{20}{c}}
		\mat J\\
		
		\\
		\mat {\phi }
		\end{array}} \right] = \left[ {\begin{array}{*{20}{c}}
		\mat w_\Lambda\\
		\\
		c
		\end{array}} \right],
	\end{equation}
	where the matrix entries are
	\begin{align*}
	{(\mat V_{\Lambda_i \Lambda_j}^v)_{mn}} &= {\left\langle {{\vv \lambda_m^{\Lambda_j}}(\vv{r}), \,\nabla {\mc V}{\vv \lambda_n^{ \Lambda_i}}(\vv{r})} \right\rangle _{\Lambda} }\,,\\
	{(\mat G_{\Lambda_i}^v)_{mn}} &= {\left\langle {{\vv \lambda_m^{\Lambda_i}}(\vv{r}), \,(\underline{\underline I} - \underline{\underline \kappa} ){\vv \lambda_n^{ \Lambda_i}}(\vv{r})} \right\rangle _{\Lambda} }\,,\\
	{(\mat D_{\Gamma_i \Lambda_j}^v)_{mn}} &= (\sigma_{i+1}-\sigma_i){\left\langle {{\vv \lambda_m^{\Lambda_j}}(\vv{r}), \,\nabla{\mc D}{\psi _n^{\Gamma_i}}(\vv{r})} \right\rangle _{\Lambda} }\,,\\
	{(\mat V^{s}_{\Lambda_i \Gamma_j})_{mn}} &= {\left\langle {{\psi _m^{\Gamma_j}}(\vv{r}), \,{\mc V} {\vv \lambda_n^{\Lambda_i}}(\vv{r})} \right\rangle _{\Gamma} }\,,\\
	{(\mat D_{\Gamma_i \Gamma_j}^{s})_{mn}} &=  (\sigma_{i+1}-\sigma_i){\left\langle {{\psi_m^{\Gamma_j}}(\vv{r}), \,{\mc D}{\psi _n^{\Gamma_i}}(\vv{r})} \right\rangle _{\Gamma} }\,, \\
	{(\mat G_{\Gamma_i}^s)_{mn}} &= \frac{{({\sigma _{i + 1}} + {\sigma _i})}}{2}{\left\langle {{\psi _m^{\Gamma_i}}(\vv{r}), \,{\psi _n^{\Gamma_i}}(\vv{r})} \right\rangle _{\Gamma} }\,,\\
	\end{align*}
	and, ${\left\langle {f,g} \right\rangle _x} = \int_x {f \cdot g} \,\diff x$ denotes the duality product. The entries of the right-hand side are
	\begin{align*}
	{({\mat w_\Lambda })_m} &= {\left\langle {{\vv{\lambda}_m}(\vv{r}), \,{\nabla v_\text{dip} }(\vv{r})} \right\rangle _{\Lambda} }\,,\\
	{({\mat {v_d}_\Gamma })_m} &= {\left\langle {{\psi_m}(\vv{r}), \,{v_\text{dip} }(\vv{r})} \right\rangle _{\Gamma}}\,.
	\end{align*}

	
	\subsection{Discretization of the single layer-wire formulation}
	Similarly to the previous approach, the surface unknown $q$ in \cref{eq:SL1,eq:SL2} is discretized with pyramid basis functions and the current density $\vv{\mathbb{J}}$ is discretized with hat basis functions. \Cref{eq:SL1,eq:SL2} are then tested with pyramid and hat functions respectively. This gives rise to the system
	\begin{equation} 
	\left[ {\begin{array}{*{20}{c}}
		{\mat G_{\Lambda _i}^v + \mat V_{{\Lambda _i}{\Lambda _j}}^v}&\vline& {\mat K_{{\Gamma _i}{\Lambda _j}}^v}\\
		\\
		{\mat W_{{\Lambda _i}{\Gamma _j}}^s}&\vline& {\mat G_{{\Gamma _i}}^s + \mat K_{{\Gamma _i}{\Gamma _j}}^s}
		\end{array}} \right]\left[ {\begin{array}{*{20}{c}}
		\mat J\\
		\\
		\mat q
		\end{array}} \right] = \left[ {\begin{array}{*{20}{c}}
		\mat w_\Lambda\\
		\\
		{{\mat{V_s}_\Gamma }}
		\end{array}} \right]
	\end{equation}
	where the matrix entries are
	\begin{align*}
	{(\mat V_{\Lambda_i \Lambda_j}^v)_{mn}} &= {\left\langle {{\vv \lambda_m^{\Lambda_j}}(\vv{r}), \,\nabla {\mc V}{\vv \lambda_n^{ \Lambda_i}}(\vv{r})} \right\rangle _{\Lambda} }\,,\\
	{(\mat G_{\Lambda_i }^v)_{mn}} &= {\left\langle {{\vv{\lambda}_m^{\Lambda_i}}(\vv{r}), \,(\underline{\underline I} - \underline{\underline \kappa} ){\vv{\lambda}_n^{\Lambda_i}}(\vv{r})} \right\rangle _{\Lambda} }\,,\\
	{(\mat K_{\Gamma_i \Lambda_j}^v)_{mn}} &= {\left\langle {{\vv \lambda_m^{\Lambda_j}}(\vv{r}), \,{\mc K}{\psi _n^{\Gamma_i}}(\vv{r})} \right\rangle _{\Lambda} }\,,\\
	{(\mat W^{s}_{\Lambda_i \Gamma_j})_{mn}} &= {\left\langle {{\psi _m^{\Gamma_j}}(\vv{r}), \,{\mc W} {\vv \lambda_n^{\Lambda_i}}(\vv{r})} \right\rangle _{\Gamma} }\,,\\
	{(\mat G_{\Gamma_i}^s)_{mn}} &= \frac{{({\sigma _{i + 1}} + {\sigma _i})}}{2({\sigma _{i + 1}} - {\sigma _i})}{\left\langle {{\psi _m^{\Gamma_i}}(\vv{r}), \,{\psi _n^{\Gamma_i}}(\vv{r})} \right\rangle _{\Gamma} }\,,\\
	{(\mat K_{\Gamma_i \Gamma_j}^{s})_{mn}} &= {\left\langle {{\psi_m^{\Gamma_j}}(\vv{r}), \,{\mc K}{\psi _n^{\Gamma_i}}(\vv{r})} \right\rangle _{\Gamma} }\,,
	\end{align*}
	and where the entries of the right-hand side are
	\begin{align*}
	{({\mat w_\Lambda })_m} &= {\left\langle {{\vv{\lambda}_m}(\vv{r}), \,{\nabla v_\text{dip} }(\vv{r})} \right\rangle _{\Lambda} }\,,\\
	{({\mat{V_s}_\Gamma })_m} &= {\left\langle {{\psi_m}(\vv{r}), \,{v_\text{s} }(\vv{r})} \right\rangle _{\Gamma}}\,.
	\end{align*}

	\subsection{Discretization of the symmetric-wire formulation}

	In contrast with the two previous approaches, the symmetric formulation (\cref{eq:Sym1,eq:Sym2,eq:Sym3}) has two surface unknowns:  the potential $\phi$ which is discretized with pyramid basis functions, the current flux with patch basis functions and the current density $\vv{\mathbb{J}}$ with hat basis functions. \Cref{eq:Sym1,eq:Sym2,eq:Sym3} are tested with patch, pyramid and hat basis functions respectively, resulting in the following system of equations
	\begin{equation}
	\left[ {\begin{array}{*{20}{c}}
		{\mat G_{\Lambda _i}^v -  \mat V_{{\Lambda _i}{\Lambda _j}}}&\vline& {\mat D_{{\Gamma _n}{\Lambda _m}}^v} &\vline& {{\mat S_{{\Gamma _n}{\Lambda _m}}^v}}\\
		\\
		{\mat W_{{\Lambda _i}{\Gamma _j}}^s} &\vline& {\mat N_{{\Gamma _i}{\Gamma _j}}^s} &\vline& {\mat K_{{\Gamma _i}{\Gamma _j}}^s}\\
		\\
		{\mat V_{{\Lambda _m}{\Gamma _n}}^s}&\vline& {\mat D_{{\Gamma _i}{\Gamma _j}}^s} &\vline& {\mat S_{{\Gamma _i}{\Gamma _j}}^s}
		\end{array}} \right]\left[ {\begin{array}{*{20}{c}}
		\mat J\\
		\\
		\mat \phi \\ \\
		\mat d
		\end{array}} \right] = \left[ {\begin{array}{*{20}{c}}
		\mat w_\Lambda \\
		\\
		{{\mat{V_s}_\Gamma }}\\ \\
		{{\mat{V_d}_\Gamma }}
		\end{array}} \right]
	\end{equation}
	in which the system entries are defined as
	\begin{align*}
	{(\mat V_{\Lambda_i \Lambda_j}^v)_{mn}} &= {\left\langle {{\vv \lambda_m^{\Lambda_j}}(\vv{r}), \,\nabla {\mc V}{\vv \lambda_n^{ \Lambda_i}}(\vv{r})} \right\rangle _{\Lambda} }\,,\\
	{(\mat G_{\Lambda_i }^v)_{mn}} &= {\left\langle {{\vv{\lambda}_m^{\Lambda_i}}(\vv{r}), \,(\underline{\underline I} - \underline{\underline \kappa} ){\vv{\lambda}_n^{\Lambda_i}}(\vv{r})} \right\rangle _{\Lambda} }\,,\\
	{(\mat D_{\Gamma_i \Lambda_j}^v)_{mn}} &= {\left\langle {{\vv \lambda_m^{\Lambda_j}}(\vv{r}), \,\theta_i {\mc D}{\psi _n^{\Gamma_i}}(\vv{r})} \right\rangle _{\Lambda} }\,,\\
	{(\mat S_{\Gamma_i \Lambda_j}^v)_{mn}} &= {\left\langle {{\vv \lambda_m^{\Lambda_j}}(\vv{r}), \,\theta_i {\mc S}{\psi _n^{\Gamma_i}}(\vv{r})} \right\rangle _{\Lambda} }\,,\\
	{(\mat W^{s}_{\Lambda_i \Gamma_j})_{mn}} &= {\left\langle {{\psi _m^{\Gamma_j}}(\vv{r}), \,\theta_j {\mc W} {\vv \lambda_n^{\Lambda_i}}(\vv{r})} \right\rangle _{\Gamma} }\,,\\
	{(\mat V^{s}_{\Lambda_i \Gamma_j})_{mn}} &= {\left\langle {{\psi _m^{\Gamma_j}}(\vv{r}), \,\theta_j {\mc V}{\vv \lambda_n^{\Lambda_i}}(\vv{r})} \right\rangle _{\Gamma} }\,,\\
	{(\mat N_{\Gamma_i \Gamma_j}^{s})_{mn}} &= {\left\langle {{\psi_m^{\Gamma_j}}(\vv{r}), \,\alpha_{ij} {\mc N}{\psi _n^{\Gamma_i}}(\vv{r})} \right\rangle _{\Gamma} }\,,\\
	{(\mat K_{\Gamma_i \Gamma_j}^{s})_{mn}} &= {\left\langle {{\psi_m^{\Gamma_j}}(\vv{r}), \,\beta_{ij} {\mc K}{\varphi _n^{\Gamma_i}}(\vv{r})} \right\rangle _{\Gamma} }\,,\\
	{(\mat D_{\Gamma_i \Gamma_j}^{s})_{mn}} &= {\left\langle {{\varphi_m^{\Gamma_j}}(\vv{r}), \,\beta_{ij}{\mc D}{\psi _n^{\Gamma_i}}(\vv{r})} \right\rangle _{\Gamma} }\,,\\
	{(\mat S_{\Gamma_i \Gamma_j}^{s})_{mn}} &= {\left\langle {{\varphi_m^{\Gamma_j}}(\vv{r}), \,\gamma_{ij} {\mc S}{\varphi _n^{\Gamma_i}}(\vv{r})} \right\rangle _{\Gamma} }\,,\\
	\end{align*}
    the entries of the right-hand side are
	\begin{align*}
	{({\mat w_{\Lambda_j} })_m} &= {\left\langle {{\vv{\lambda}_m^{\Lambda_j}}(\vv{r}), \,{\nabla v_\text{dip} }(\vv{r})} \right\rangle _{\Lambda} }\,,\\
	{({\mat {v_d}_{\Gamma_j} })_m} &= {\left\langle {{\psi_m^{\Gamma_j}}(\vv{r}), \,\theta_j{v_\text{dip} }(\vv{r})} \right\rangle _{\Gamma}}\,,\\
	{({\mat{V_s}_{\Gamma_j} })_m} &= {\left\langle {{\psi_m^{\Gamma_j}}(\vv{r}), \,\theta_j {v_\text{s} }(\vv{r})} \right\rangle _{\Gamma}}\,,
	\end{align*}
	%
	and the coefficients $\alpha$, $\beta$, $\gamma$ and $\theta$ are defined in \Cref{tab:table1}.
	
	\begin{table}
		\centering
		\begin{tabular}{|c|c|c|c|c|}
			\hline
			Condition & $\alpha_{ij}$       & $\beta_{ij}$ & $\gamma_{ij}$ & $\theta_j$ \\ \hline
			$j=1$     & $-$                 & $-$          & $-$                           &  1  \\
			$j=i$     & $\sigma_i+\sigma_j$ & $-2$         & $\sigma_i^{-1}+\sigma_j^{-1}$ & $-$ \\ 
			$j=i-1$   & $-\sigma_i$         & $1$          & $-\sigma_i^{-1}$              & $-$ \\ 
			$j=i+1$   & $-\sigma_j$         & $1$          & $-\sigma_j^{-1}$              & $-$ \\ 
			otherwise      & $0$                 & $0$          & 0                             & 0   \\
			\hline
		\end{tabular}
		\caption{Definition of the different coefficients associated with the discretization of the symmetric formulation.}
		\label{tab:table1}
	\end{table}
		
\section{Numerical results} \label{sec:numerical_results}
	In this section the newly developed integral formulations are validated and their performances are studied through several numerical examples. The parameters of the simulations are given in normalized units.

\subsection{Convergence of the solution}

\begin{figure}
	\centering
    \begin{subfigure}[t]{0.49\textwidth}%
    \centering
	\includegraphics [trim=3.1cm 3.0cm 3.0cm 2.3cm, clip=true, totalheight=0.19\textheight]{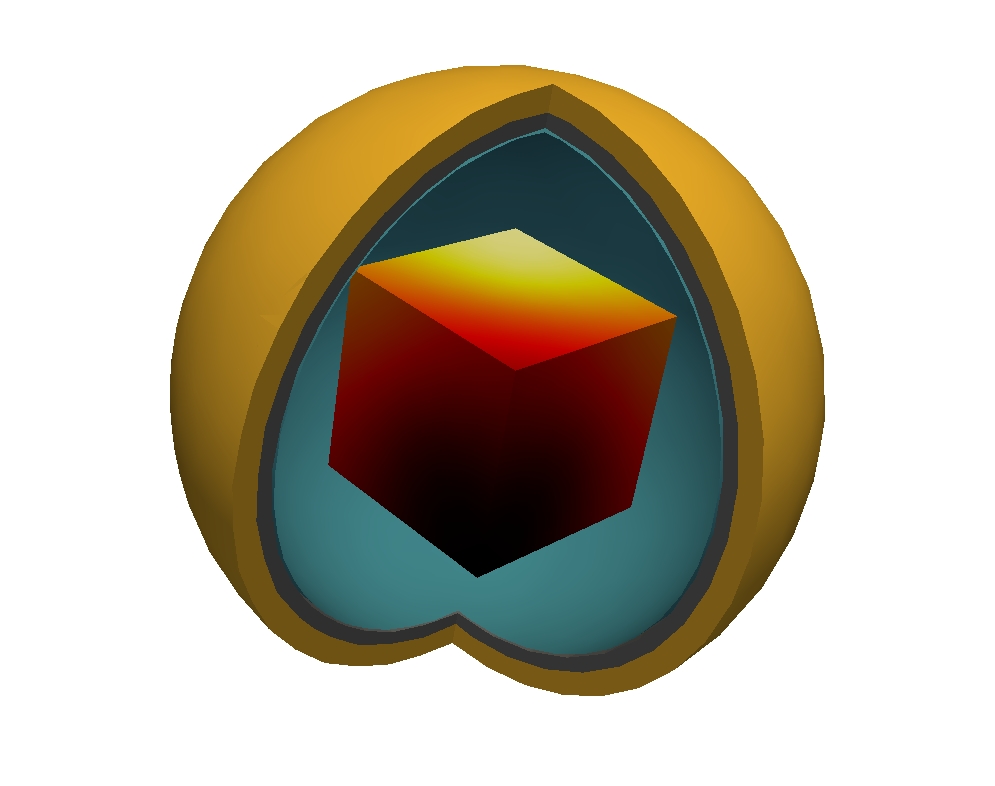}
	\caption{ \label{subfig:Reference_Model}}
	\end{subfigure}
    \begin{subfigure}[t]{0.49\textwidth}%
    \centering
	\includegraphics [trim=3.7cm 5.0cm 19.0cm 3.5cm, clip=true, totalheight=0.2\textheight]{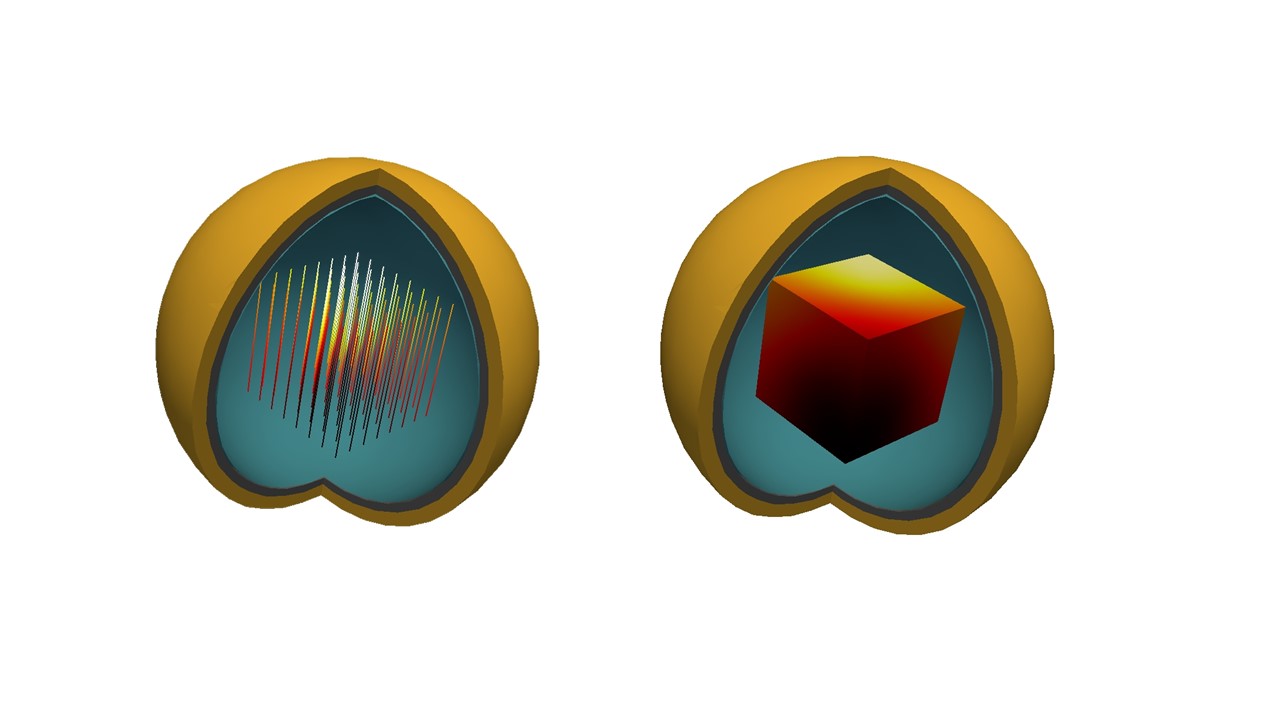}
	\subcaption{ \label{subfig:Simulated_Model}}
	\end{subfigure}
	\caption{Anisotropic cube inside a three layered sphere: (\subref{subfig:Reference_Model}) reference model and (\subref{subfig:Simulated_Model}) simulated model.}
	\label{fig:Cube3Sphers_Model}
\end{figure}

In order to demonstrate that the proposed formulations are capable of capturing the anisotropic conductivity caused by the brain fibers and do converge to the exact solution, we have simulated a cubic block whose anisotropic conductivity is $\num{10}$ along the $z$ axis and $\num{1}$ in the $(x,\,y)$ plane, residing inside a three layered sphere (\Cref{subfig:Reference_Model}). The radii of the spheres are $\num{0.87}, \num{0.92}$ and $\num{1}$ respectively. The cube, whose side length is equal to $\num{0.7}$, is placed at their center. The conductivity of the different spheres are $\num{1}$, $\num{1/15}$ and $\num{1}$ respectively. A current source with a dipolar moment equal to $[\num{1},\num{1},\num{1}]$ is set at $[\num{0.4}, \num{0}, \num{0}]$. In order to account for the anisotropic effect of the cube with our formulations, we have created a grid of $64$ equally spaced fiber, as illustrated in \cref{subfig:Simulated_Model}. The wires have a radius of $0.05$ and their conductivity is set to be $10$ along the wires and $1$ in their transverse direction. A convergence analysis has been carried out in which the model is discretized with increasingly refined mesh (the number of wires has been kept constant). Note that the exact wire structure shown in \Cref{subfig:Simulated_Model} could have been solved with FEM and used as reference rather than the one obtained with the cube structure in \Cref{subfig:Reference_Model}, however a more extreme case was preferred to illustrate the merits of the new schemes by choosing a different modelization of the underlying physics.
%
%

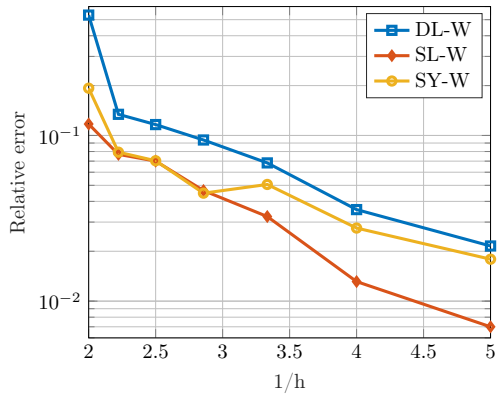
\begin{figure}
	\centering
    \resizebox {0.5\textwidth} {!} {
%
%
\definecolor{mycolor1}{rgb}{0.00000,0.44700,0.74100}%
\definecolor{mycolor2}{rgb}{0.85000,0.32500,0.09800}%
\definecolor{mycolor3}{rgb}{0.92900,0.69400,0.12500}%
\begin{tikzpicture}

\begin{axis}[%
xmin=2,
xmax=5,
xlabel style={font=\color{white!15!black}},
xlabel={1/h},
ymode=log,
ymin=0.006,
ymax=.6,
yminorticks=true,
ylabel style={font=\color{white!15!black}},
ylabel={Relative error},
axis background/.style={fill=white},
xmajorgrids,
ymajorgrids,
yminorgrids,
legend style={legend cell align=left, align=left, draw=white!15!black}
]
\addplot [color=mycolor1, line width=1.5pt, mark=square, mark options={solid, mycolor1}]
  table[row sep=crcr]{%
2	0.532869\\
2.22222222222222	0.13417\\
2.5	0.116337\\
2.85714285714286	0.0939\\
3.33333333333333	0.06829\\
4	0.03569\\
5	0.02152032\\
};
\addlegendentry{DL-W}

\addplot [color=mycolor2, line width=1.5pt, mark=diamond, mark options={solid, mycolor2}]
  table[row sep=crcr]{%
2	0.1172354\\
2.22222222222222	0.076914\\
2.5	0.07004\\
2.85714285714286	0.0465\\
3.33333333333333	0.032405\\
4	0.013122\\
5	0.007\\
};
\addlegendentry{SL-W}

\addplot [color=mycolor3, line width=1.5pt, mark=o, mark options={solid, mycolor3}]
  table[row sep=crcr]{%
2	0.192891291\\
2.22222222222222	0.07927726\\
2.5	0.070374\\
2.85714285714286	0.0448365\\
3.33333333333333	0.05061\\
4	0.02761\\
5	0.01791\\
};
\addlegendentry{SY-W}

\end{axis}
\end{tikzpicture}%
    }
	\caption{The relative error of the DLW, SLW and SYW formulations as a function of the average edge length which shows the convergence of the solutions to the reference solution obtained with FEM. The simulated geometry is illustrated in \Cref{fig:Cube3Sphers_Model}.}
	\label{fig:Convergence_Plot}
\end{figure}

\Cref{fig:Convergence_Plot} reports the obtained relative error as a function of the mesh edge length; a FEM solution corresponding to highly refined mesh is used as reference. It is clear that the three formulations converge to the reference solution and can indeed account for the anisotropy of the medium.

\subsection{Accuracy for different dipole eccentricities}

\begin{figure}
	\centering
	\includegraphics[width=0.6\linewidth]{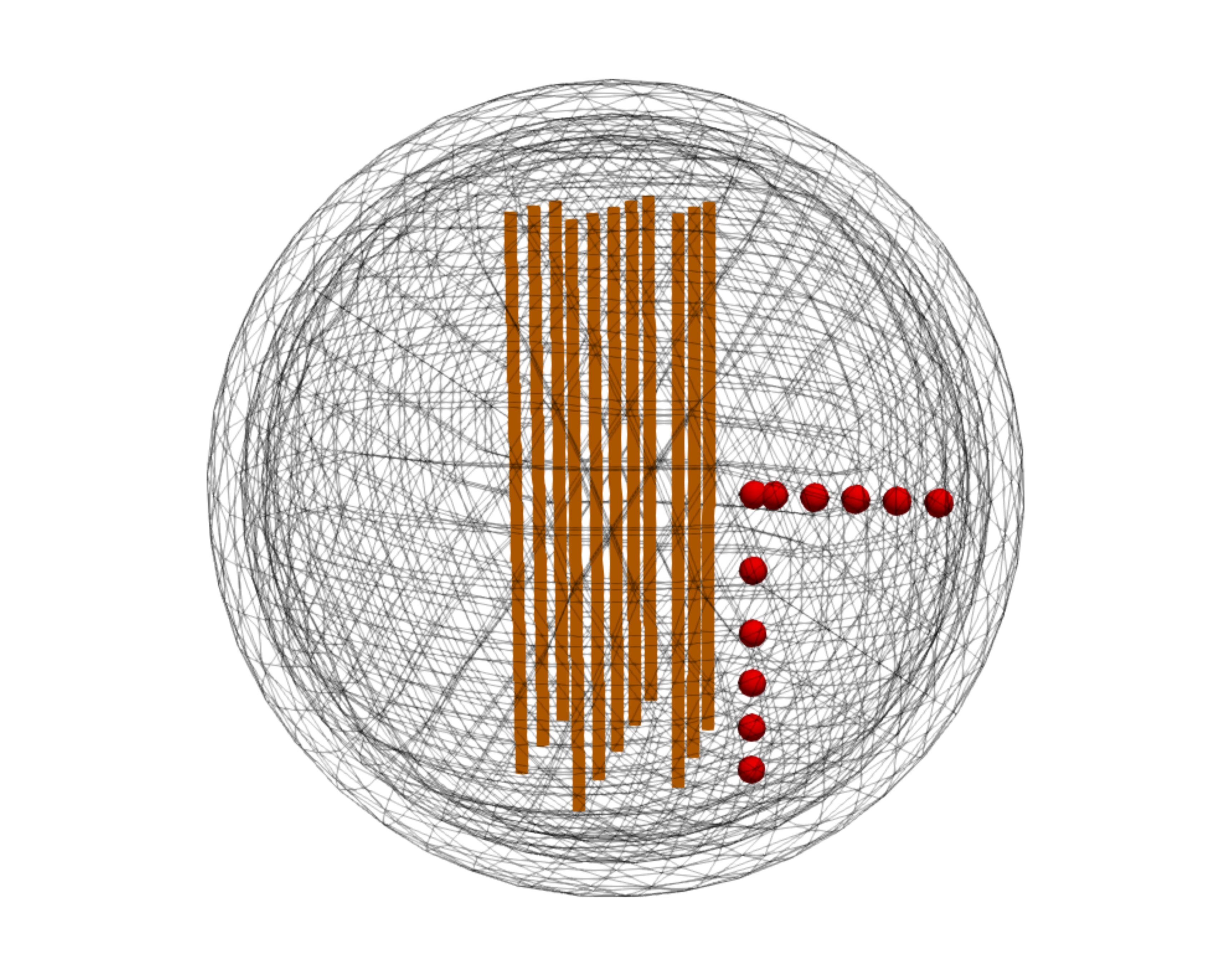}
	\caption{Geometry of the simulated model; the red dots indicate the position of the current sources.}
	\label{fig:threesphere12wires}
\end{figure}

\begin{figure}
	\centering
    \begin{subfigure}[t]{0.49\textwidth}%
    \resizebox {\textwidth} {!} {
%
%
\definecolor{mycolor1}{rgb}{0.00000,0.44700,0.74100}%
\definecolor{mycolor2}{rgb}{0.85000,0.32500,0.09800}%
\definecolor{mycolor3}{rgb}{0.92900,0.69400,0.12500}%
\definecolor{mycolor4}{rgb}{0.49400,0.18400,0.55600}%
\begin{tikzpicture}

\begin{axis}[%
xmin=0.3,
xmax=0.8,
xlabel style={font=\color{white!15!black}},
xlabel={Source Eccentricity},
ymode=log,
ymin=0.003,
ymax=0.5,
yminorticks=true,
ylabel style={font=\color{white!15!black}},
ylabel={Relative Error},
axis background/.style={fill=white},
xmajorgrids,
ymajorgrids,
yminorgrids,
legend style={legend cell align=left, align=left, draw=white!15!black, anchor=west, at={(0.03,0.5)}},
]
\addplot [color=mycolor1, line width=1.5pt, mark=triangle, mark options={solid, mycolor1}]
  table[row sep=crcr]{%
0.3	0.175485\\
0.4	0.156363877\\
0.5	0.1401595\\
0.6	0.13775\\
0.7	0.15046\\
0.8	0.172873\\
};
\addlegendentry{AN}

\addplot [color=mycolor2, line width=1.5pt, mark=square, mark options={solid, mycolor2}]
  table[row sep=crcr]{%
0.3	0.00997\\
0.4	0.011127\\
0.5	0.013437\\
0.6	0.01775\\
0.7	0.029543\\
0.8	0.069111\\
};
\addlegendentry{DL-W}

\addplot [color=mycolor3, line width=1.5pt, mark=diamond, mark options={solid, mycolor3}]
  table[row sep=crcr]{%
0.3	0.00396744\\
0.4	0.00403\\
0.5	0.005601\\
0.6	0.00741\\
0.7	0.01044543\\
0.8	0.022106\\
};
\addlegendentry{SL-W}

\addplot [color=mycolor4, line width=1.5pt, mark=o, mark options={solid, mycolor4}]
  table[row sep=crcr]{%
0.3	0.0097974\\
0.4	0.011008\\
0.5	0.013506\\
0.6	0.01731534\\
0.7	0.01983\\
0.8	0.039222\\
};
\addlegendentry{SY-W}

\end{axis}
\end{tikzpicture}%
	}
	\caption{ \label{subfig:RE_Ecc_Along}}
	\end{subfigure}
\begin{subfigure}[t]{0.49\textwidth}%
    \resizebox {\textwidth} {!} {
%
%
\definecolor{mycolor1}{rgb}{0.00000,0.44700,0.74100}%
\definecolor{mycolor2}{rgb}{0.85000,0.32500,0.09800}%
\definecolor{mycolor3}{rgb}{0.92900,0.69400,0.12500}%
\definecolor{mycolor4}{rgb}{0.49400,0.18400,0.55600}%
\begin{tikzpicture}

\begin{axis}[%
xmin=0.3,
xmax=0.8,
xlabel style={font=\color{white!15!black}},
xlabel={Source Eccentricity},
ymode=log,
ymin=0.003,
ymax=0.4,
yminorticks=true,
ylabel style={font=\color{white!15!black}},
ylabel={Relative Error},
axis background/.style={fill=white},
xmajorgrids,
ymajorgrids,
yminorgrids,
legend style={legend cell align=left, align=left, draw=white!15!black, anchor=west, at={(0.03,0.5)}}
]
\addplot [color=mycolor1, line width=1.5pt, mark=triangle, mark options={solid, mycolor1}]
  table[row sep=crcr]{%
0.3	0.175485\\
0.4	0.13728\\
0.5	0.109295\\
0.6	0.086598\\
0.7	0.067499\\
0.8	0.0506\\
};
\addlegendentry{AN}

\addplot [color=mycolor2, line width=1.5pt, mark=square, mark options={solid, mycolor2}]
  table[row sep=crcr]{%
0.3	0.009771\\
0.4	0.01148\\
0.5	0.014\\
0.6	0.01848\\
0.7	0.02862\\
0.8	0.072734\\
};
\addlegendentry{DL-W}

\addplot [color=mycolor3, line width=1.5pt
, mark=diamond, mark options={solid, mycolor3}]
  table[row sep=crcr]{%
0.3	0.003954\\
0.4	0.003767\\
0.5	0.00414\\
0.6	0.0050501\\
0.7	0.00768\\
0.8	0.02103\\
};
\addlegendentry{SL-W}

\addplot [color=mycolor4, line width=1.5pt, mark=o, mark options={solid, mycolor4} ]
  table[row sep=crcr]{%
0.3	0.009758\\
0.4	0.01118\\
0.5	0.013071\\
0.6	0.01594\\
0.7	0.02076\\
0.8	0.05\\
};
\addlegendentry{SY-W}

\end{axis}
\end{tikzpicture}%
	}
	\subcaption{ \label{subfig:RE_Ecc_Trans}}
	\end{subfigure}
	\caption{Relative error of the different formulations as a function of dipole eccentricity: (\subref{subfig:RE_Ecc_Along}) along the fibers and (\subref{subfig:RE_Ecc_Trans}) away from the fibers. The simulated geometry as well as the dipole sources are shown in \Cref{fig:threesphere12wires} where a refined FEM solution was used as a reference. In the legend, AN refers to the analytical solution of the corresponding spherical geometry in the absence of the fibers.}
	\label{fig:RE_Ecc}
\end{figure}
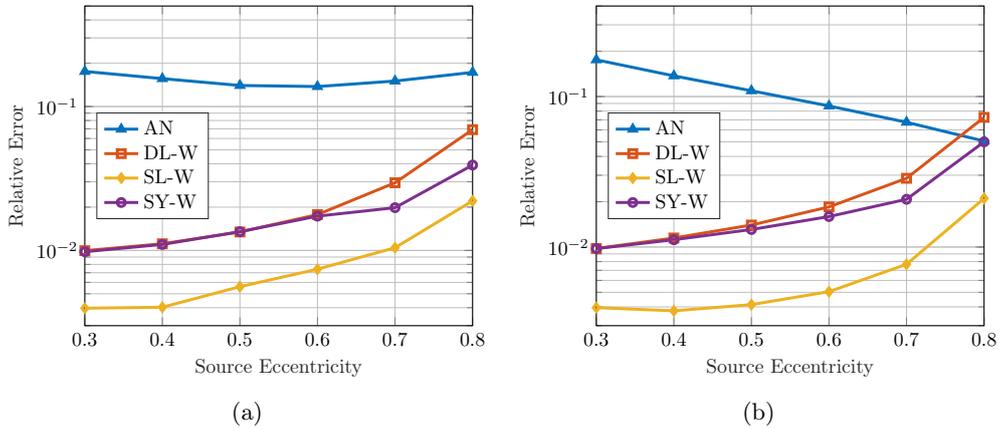

In the second test, we have studied the effect of source eccentricity on the computed potential. Three concentric spheres of radius $\num{0.87}$, $\num{0.92}$ and $\num{1}$ have been considered. Eleven vertical fibers of  radius $0.05$ were set at the coordinates summarized in \Cref{tab:table2}.
\begin{table}
    \centering
	\begin{tabular}{|l|c|c|c|c|c|c|c|c|c|c|c|}
		\hline
		x & -0.4 & -0.2 & 0 & 0.2 & 0.4 & -0.2 & 0    & 0.2  & -0.2 & 0   & 0.2 \\ \hline
		y & 0    & 0    & 0 & 0   & 0   & -0.2 & -0.2 & -0.2 & 0.2  & 0.2 & 0.2 \\ \hline
	\end{tabular}
	\caption{Coordinates of the wires in the $xy$ plane corresponding to \Cref{fig:threesphere12wires}.}
	\label{tab:table2}
\end{table} 
The conductivities of the different layers of the sphere were set to  $\num{1}$, $\num{1/15}$ and $\num{1}$ and the fibers to $10$ along the $z$ direction and $1$ in the transversal direction. The model was discretized with $642$ nodes per surface and $15$ segments per fiber. The forward problem was then solved for a varying dipole position: along and away from the fibers as shown in \cref{fig:threesphere12wires} with red dots. The computed relative error, where a high resolution FEM was used as a reference, is shown in  \cref{subfig:RE_Ecc_Trans,subfig:RE_Ecc_Along} for the two cases. In order to illustrate the error introduced when neglecting the anisotropic conductivity of the fibers, we have also included the relative error produced by the analytic solution of the same spherical model in the absence of the fibers. 
 	
In general the accuracy of the three numerical solutions decreases for shallow sources. This behavior is due to the singularity of the source and is in agreement with what has been reported in the literature \cite{vorwerk2017mixed, kybic2005common, rahmouni2017mixed}. It is also observed that not accounting for the anisotropic conductivity of the fibers leads to higher errors, especially in vicinity of the fibers. As expected, these errors decrease when the source is moved away from the fibers (\cref{subfig:RE_Ecc_Trans}) and remains stable when moving in their vicinity (\cref{subfig:RE_Ecc_Along}).



\subsection{Application to a realistic head mesh}

\begin{figure}
	\centering
	\begin{subfigure}[t]{0.51\textwidth}
	\includegraphics[width=\textwidth]{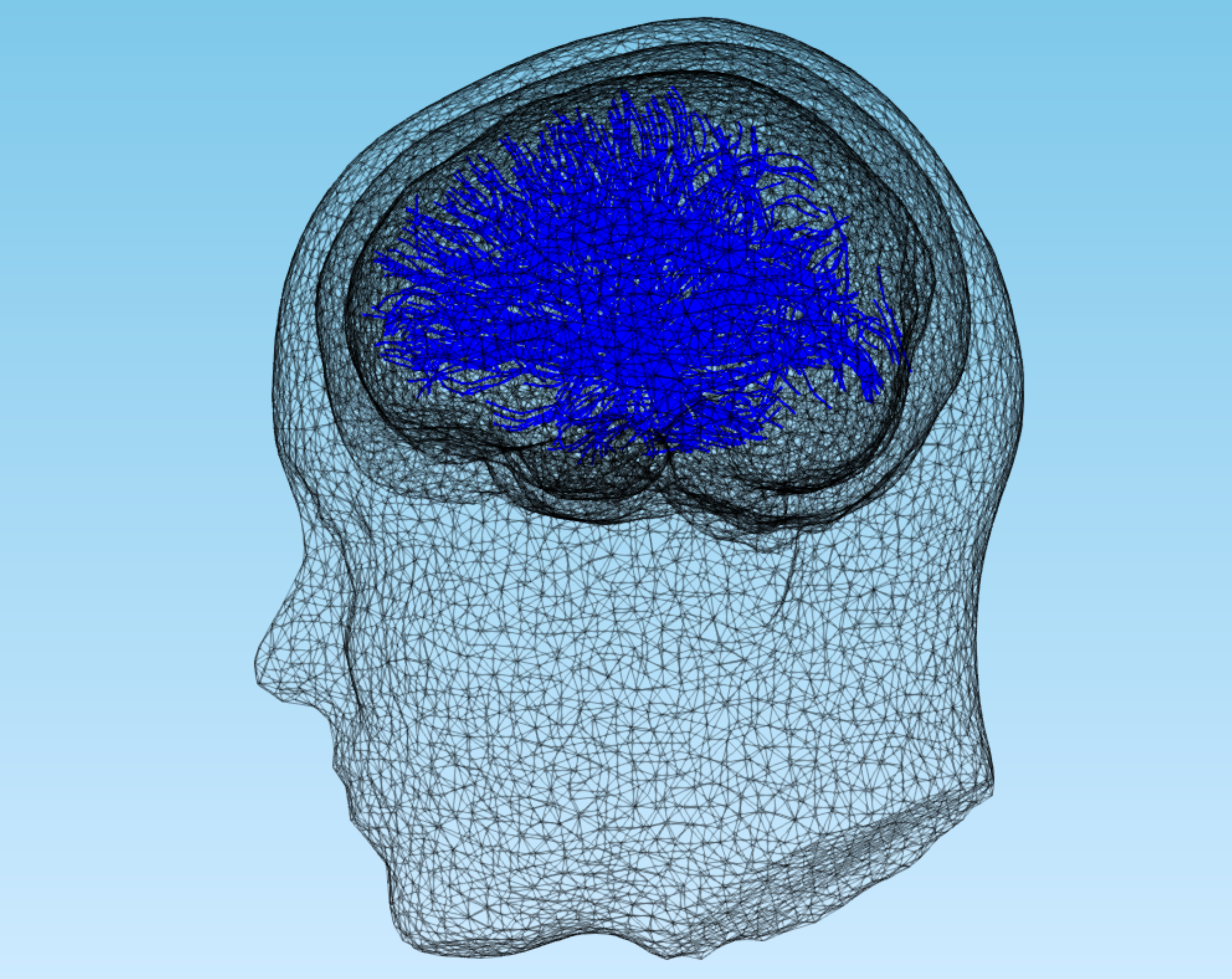}
	\caption{\label{subfig:realisticheadmodel}}
	
	\end{subfigure}
    \begin{subfigure}[t]{0.49\textwidth}
	\centering
	\includegraphics[width=\linewidth]{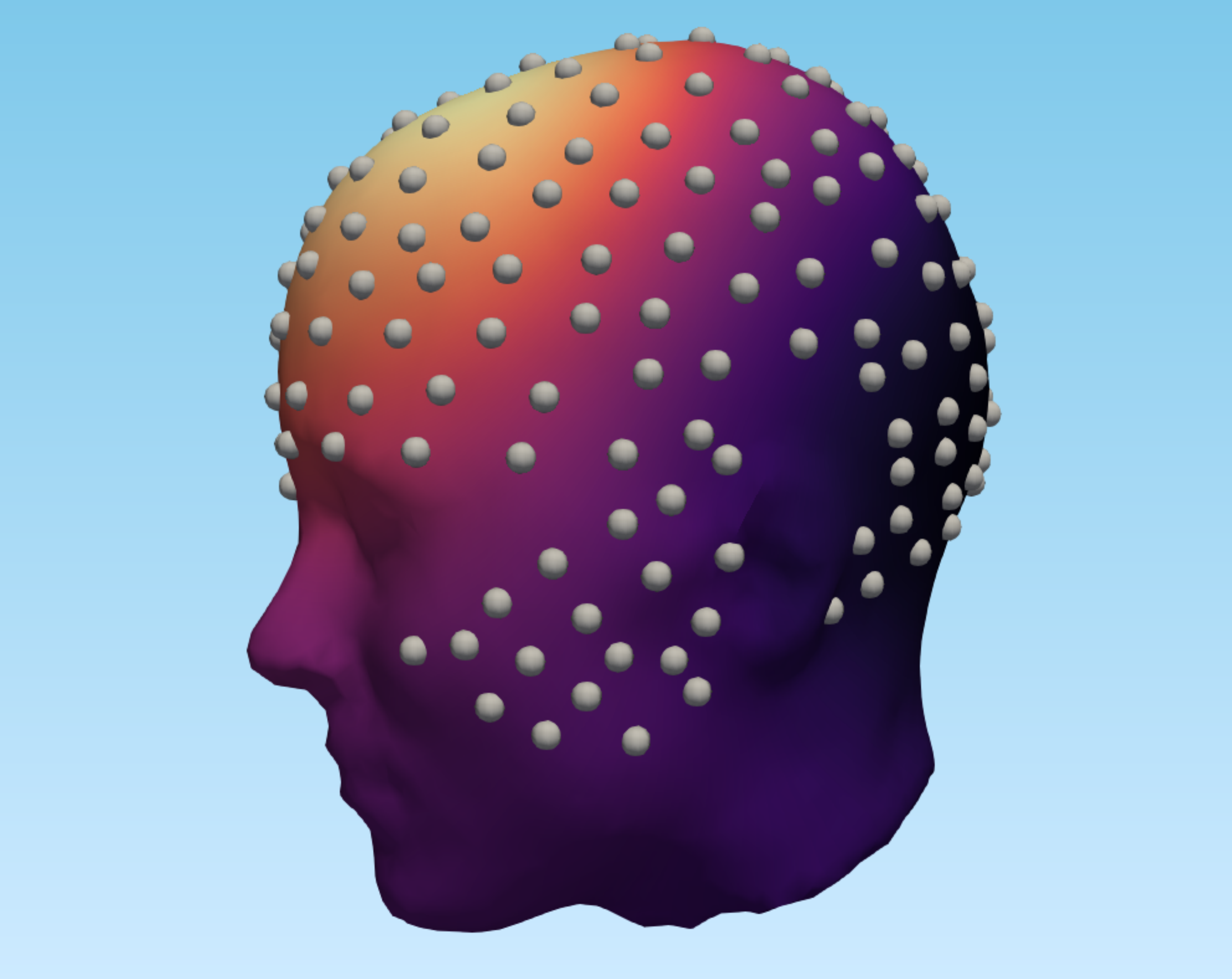}
	\caption{\label{subfig:Scalp_Potential_Electrodes}}
	
    \end{subfigure}
    \begin{subfigure}[t]{0.49\textwidth}
	\centering
	\includegraphics[width=\linewidth]{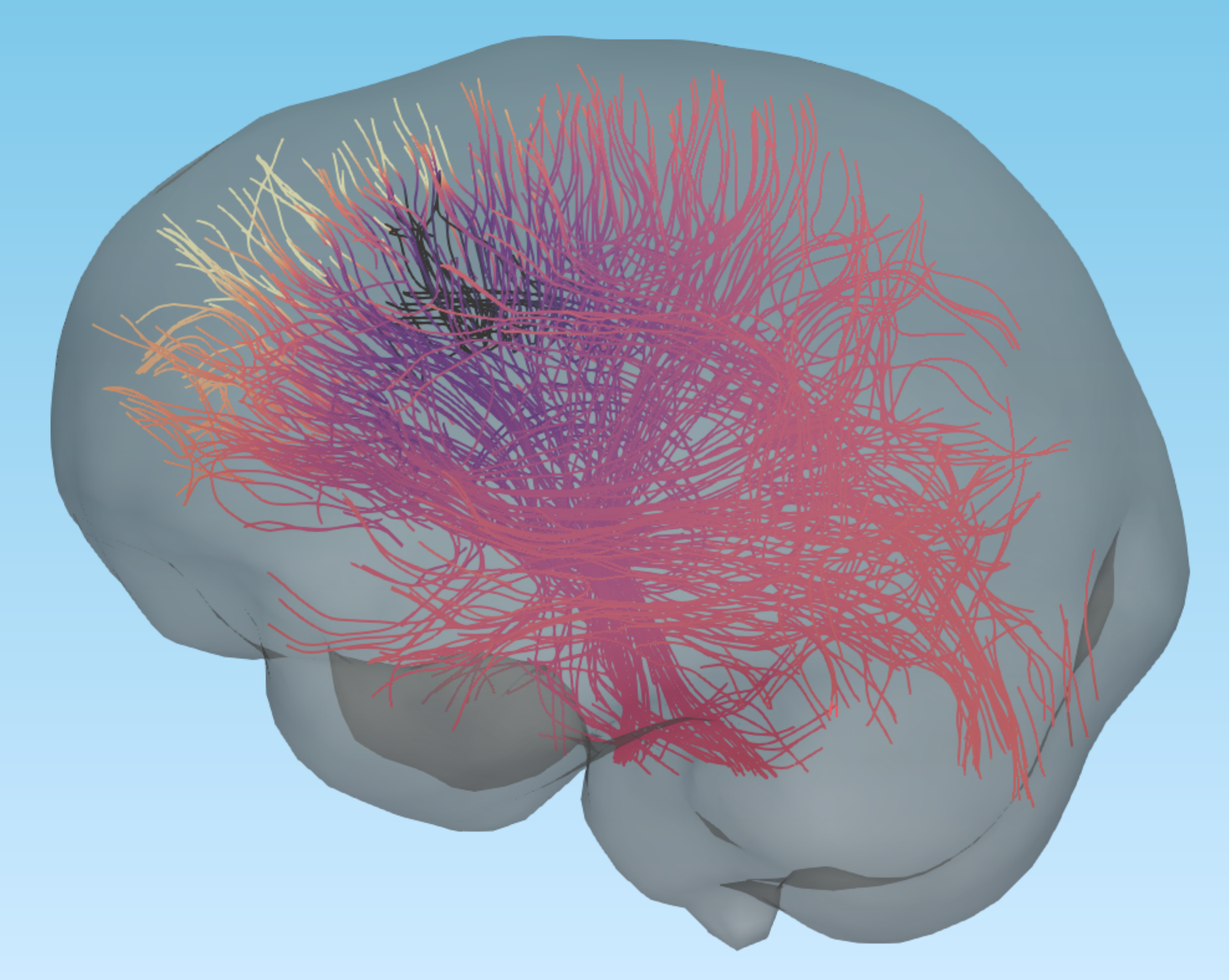}
	\caption{\label{subfig:Brain_Fiber_Current_Density}}
	
    \end{subfigure}
    \caption{MRI-based head model: (\subref{subfig:realisticheadmodel}) the simulated head model composed of white matter fibers, brain, skull and scalp surfaces; (\subref{subfig:Scalp_Potential_Electrodes}) computed distribution of the electric potential on the scalp, the dots indicate the position of electrodes of a high density EEG device; (\subref{subfig:Brain_Fiber_Current_Density}) magnitude of the calculated current density along the fibers.}
    \label{fig:models}
\end{figure}

\begin{figure}
	\centering
	\resizebox {0.7\textwidth} {!} {
	\input {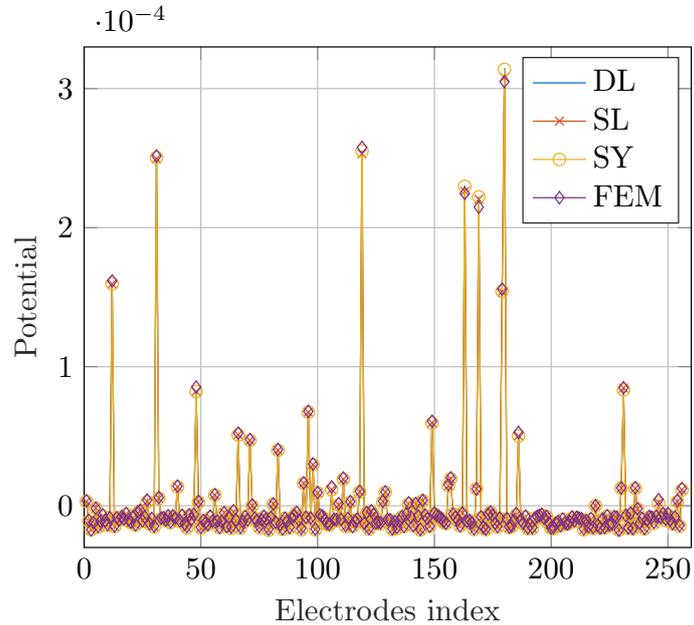}
	}
	\caption{Electric potential computed at electrodes position shown in \Cref{subfig:Scalp_Potential_Electrodes}. }
	\label{fig:Electrode_Potential}
\end{figure}

As a last numerical test, a realistic head model obtained from MRI images is considered (\Cref{subfig:realisticheadmodel}). Using standard procedures (see for example \cite{windhoff2013electric}), we have constructed a 3 layered mesh in which each domain represents the brain, the skull and the scalp, each of which is made of $\num{6248}$, $\num{8328}$ and $\num{9346}$ triangles, respectively. Furthermore, the white matter fibers are recovered using DTI-based tractography implemented in \cite{toussaint2007medinria}. The conductivity of the different tissues is set to $0.33$, $0.067$ and $0.33$ for the scalp, skull and brain respectively. The conductivity of the fibers is set to be $0.33$ in their local transverse direction and $10$ times greater in their longitudinal direction. Following the EGI system \cite{luu2005determination}, a set of $256$ electrodes has been placed on the scalp as shown in \Cref{subfig:Scalp_Potential_Electrodes}. At these positions, the electric potential was computed using the newly introduced schemes. For the sake of comparison, we have also computed the solution with FEM, on a volume mesh of 10 million tetrahedrons. We show the results obtained in \Cref{fig:Electrode_Potential}, where we observe that the four formulations are in agreement. In \Cref{subfig:Brain_Fiber_Current_Density}, we plot the magnitude of the current density along the fibers.

\section{Conclusion} \label{sec:conclusion}

The correct modeling of the electric properties of the head is crucial for an accurate forward solution and, consequently, for brain source reconstruction. This includes the anisotropic behavior of the white matter, given its impact on the scalp potential. In this paper, we have presented new integral techniques that can handle the anisotropic conductivity profile of the head and thus extend the application of conventional BEM approaches. The one dimensional nature of the wire basis functions ensure the computational efficiency of the schemes. It has been shown throughout several numerical tests that the computed potential exhibit high accuracy and stability making it a competitive alternative to differential equations based methods.

\section*{Acknowledgment}

This work has been funded in part by the European Research Council (ERC) under the European Union's Horizon 2020 research and innovation program (ERC project 321, grant No. 724846).


\bibliography{Bibliography}

\end{document}